\documentstyle[spie,psfig]{article} 

\title{A simulation code to assist designing space missions of the
Airwatch type} 

\author{T. Montaruli\supit{a}, R. Bellotti\supit{b}, F. Cafagna\supit{a}, 
M. Circella\supit{a}, C. N. De Marzo\supit{b} and P. Lipari\supit{c} 
\skiplinehalf 
\supit{a} Sezione INFN di Bari, Via Amendola 173, 70126 Bari, Italy 
\skiplinehalf 
\supit{b} Dip. di Fisica, Univ. di Bari and Sezione INFN di 
Bari, Via Amendola 173, 70126 Bari, Italy
\skiplinehalf 
\supit{c} Sezione INFN di Roma, Univ. di Roma I, P. A. Moro, 00185 
Roma, Italy
}

\authorinfo{Further author information:\\ 
(Send correspondence to)\\
C.~N.~D.: E-mail: demarzo@ba.infn.it\\ 
T.~M.: E-mail: montaruli@ba.infn.it\\
}

\begin{document} 
\maketitle 

\begin{abstract}
The design of an Airwatch type space mission can greatly
benefit from a flexible simulation code for establishing the
values of the main parameters of the experiment.
We present here a code written for this purpose. The cosmic ray primary
spectrum at very high energies, the atmosphere modelling, the 
fluorescence yield,
the photon propagation and the detector response are taken into account
in order to optimize the fundamental design parameters of the
experiment, namely orbit height, 
field of view, mirror radius, number of pixels of the focal plane, 
threshold of photo-detection.
The optimization criterion will be to maximize counting rates
versus mission cost, which imposes limits both on weight and power consumption.
Preliminary results on signals with changing
energy and zenith angle of incident 
particles are shown.
\end{abstract}



\section{INTRODUCTION}
\label{sect:intro}  

In recent years, the extremely high energy region of the cosmic ray spectrum 
($10^{18}-10^{21}$ eV) has been attracting more and more 
attention because of the
fundamental questions (in High Energy Astrophysics)
concerning the origin and propagation of detected 
particles of such high energy.
High energy cosmic rays are detected by 
measuring the large showers they 
produce in the Earth atmosphere 
which acts as a giant detector for incoming extraterrestrial
radiation. Extreme Energy Cosmic Ray (EECR) induced showers  
produce fluorescence light in the atmospheric nitrogen  
that can be observed by detectors like Fly's Eye
\cite{Baltrusaitis}. In this way the observable quantities that
can be measured in order to study the EECR are longitudinal shower
profile, depth and size of the shower maximum and arrival direction.
From these observables angular anisotropy, energy, nature and composition of
the primary particles have to be derived.

Recently the possibility of observing the fluorescence light induced by the 
EECR in the atmospheric nitrogen from space platforms 
has been proposed \cite{Linsley,Linsley1,Takahashi}, 
hence providing an alternative approach
to very large ground based arrays.
In fact, in measuring the cosmic ray spectrum up to the highest energies
severe conditions are posed by the particle flux,
which at energies higher than 3 EeV appear to decrease as $E^{-\alpha}$,
with differential spectral index $\alpha = 2.7$. 
As a consequence, observations at these energies have to cope
with fluxes lower than one particle per square kilometer per millennium.
Indeed detectors on space platforms appear to provide
the possibility of extending the EECR measurements beyond $10^{21}$ eV 
\cite{OWL,Ormes} due to the large atmospheric volume that can be looked at
by a single detector. Anyway,  
the detection of the EECRs through the fluorescence light induced 
in the atmosphere requires detectors with good sensitivity in the UV
and in the optical ranges, large collecting optics and a focal plane
segmentation up to a million pixels.
Such requirements are particularly compelling for an Airwatch type
detector \cite{Scarsi}, 
since in this case the showers will be reconstructed on the basis
of the image detected by a single satellite with proper time information.
The construction of such a detector in the present situation requires 
both a substantial R\&D and a design optimization
effort. To this aim the study of the mission configuration
with a simulation code offers a cheap and flexible way to
decide optimal values for parameters before starting the detector 
construction. 

We present here the first steps we have implemented in the
simulation of a mission of the Airwatch type.
\section{THE SIMULATION CODE ARCHITECTURE} 

The aims of the simulation code are:
\begin{itemize}
\item to study at a first order precision the feasibility of
an Airwatch type mission and its main features, such as
geometrical parameters, orbit height, energy threshold, capabilities of
the photodetector;
\item to estimate counting rates for given detector features. 
\end{itemize}
In order to fulfil
these aims many parameters are input to the algorithm and hence
are easily changeable. 

The flow-chart of the algorithm is:
\begin{enumerate}
\item geometry initialization. At this stage 3 input parameters are
required: the orbit height $H$, half of the field of view angle 
$\theta_{FOV}$, the threshold for the detection of photons in terms of
photoelectrons $(p.e.)_{thr}$. We have assumed that 
the photodetectors are capable of resolving one single p.e.;
\item initialization of the energy distribution of primaries. There are
two possible choices to run the program: using
a fixed value of the primary energy or an energy randomly generated 
from a given 
energy spectrum. In the former case the input to the
simulation is the energy of the primary, in the latter it is the minimum
energy of the spectrum $E_{min}$.
We have assumed a pure proton component 
with a structure at $E^{*} = 3.162$ EeV 
(according to Fly's Eye observations \cite{Fly}). 
The differential spectral index
of the flux is $\alpha_{1} = 3.27$ for energies lower than $E^{*}$ and
$\alpha_{2} = 2.71$ for energies higher that $E^{*}$;
\item initialization of the shower development process: initializes hadron
interaction cross
sections (according to the SIBYLL model \cite{SIBYLL}), 
electromagnetic cross-sections
(pair production and bremsstrahlung), the muon energy loss for
ionization in air. We point out that our aim here is to investigate the
experimental configuration rather than to discuss the models 
of shower development in the atmosphere at such high energy 
(above 10$^{19} eV$).
The structure of the code is such that any parametrization
of cross sections from any model can be easily implemented;
\item event generation (input: the number of events to be generated):
\begin{itemize}
\item generation of the impact point of the trajectory 
on the Earth surface;
\item generation of the direction of the track of the primary;
\item generation of the energy of the event according to the initialization
choice explained above; 
\item structure of the atmosphere and conversion between slanted depth
along the track and vertical height; 
\item shower development \footnote{At the moment the shower is initiated only 
by protons; in the future the superposition principle will be applied 
in order to consider showers initiated by heavier nuclei.}: 
the charged particles developed in the atmosphere (size) 
are counted in the proper number of steps in which the track is divided.
The secondary particles
of the shower are assumed to be in the same direction of the incident particle
in an unidimensional approximation;
\item {\it i)} fluorescence light generation; 
{\it ii)} propagation onto the light collector and focalization on the 
detector pixel matrix; {\it iii)} quantum conversion into photoelectrons; 
\item record of the relevant information for each event. 
\end{itemize}
\end{enumerate} 

In Fig.~\ref{fig1} the parameters
used for describing the region of the atmosphere seen by the detector
are schematically shown. The orbit height $H$ and $\theta_{FOV}$, together
with the radius of the Earth $R_{\oplus}$, determine the opening
angle $\theta^{*}$ of the cone with the vertex in the center of the Earth
which sees the same area $A$ as seen by the detector:

\begin{equation}
\label{eq:thetastar}
\cos\theta^{*} = \frac{(1 + \frac{H}{R_{\oplus}})\tan^{2}\theta_{FOV} + 
\sqrt{1 + \tan^{2}\theta_{FOV} - (1 + \frac{H}{R_{\oplus}})^{2}\tan^{2}
\theta_{FOV}}}{(1 + \tan^{2}\theta_{FOV})} \, .
\end{equation}

Hence, considering the reasonable values $H = 500$ km, $\theta_{FOV} = 
30^{\circ}$, then $\theta^{*} = 2.6^{\circ}$, the area on the curved
Earth surface
seen by the detector is $A = 2 \pi R^{2}_{\oplus} (1-\cos\theta^{*}) 
= 2.7 \times 10^{5}$ km$^{2}$ having a radius of 289 km.
Given this area, the acceptance is approximately
$Acc = 2 \pi A$ (because the Earth blocks cosmic rays for half of the sky)
and hence the expected rate for the integral flux considered (from Fly's
Eye data) is for $E_{min} =  10^{2}$ EeV, without
considering any efficiency:

\begin{equation}
\label{eq:acc}
Rate = Acc (km^{2} sr) \times Flux(E \ge E_{min})(km^{-2} yr^{-1} sr^{-1})
=  1.5 \times 10^{4} events/year \, .
\end{equation}

\begin{figure}[t]
   \begin{center}
   \begin{tabular}{cc}
\psfig{figure=app.ps,width=7.cm,height=7.cm}&
\psfig{figure=exp.ps,width=7.cm,height=7.cm}
   \end{tabular}
   \end{center}
   \caption[Schematical view] 


{ \label{fig1}	  

Schematical view of the detector (B). The orbit height is 
$\overline{OB}$, the radius of the Earth is $\overline{AO}$, 
the angle $\theta^{*}$ is $O\widehat{A}C$ and the angle $\theta_{FOV}$
is $O\widehat{B}C$. The plot on the right is an expanded view of 
the one on the left where an example of generated zenith angle $\theta$
is shown.} 

\end{figure} 

The geometrical parameters relevant for the detector are
the area of the mirror which collects the fluorescence photons, 
assumed 5 m$^{2}$, and the structure of the pixel matrix.
The order of magnitude of the number of pixels needed depends on
$\theta_{FOV}$, on $H$, on the length of tracks $\ell$ in the atmosphere
and on the number of sampling points along the track $N$. 
For typical values of
$\ell = 10$ km and $N = 30$: \cite{OWL}

\begin{equation}
\label{eq:pixel}
N. \; of \; pixels = \sin^{2}\theta_{FOV}N^{2} \left(\frac{2H}{\ell}\right) 
\sim 10^{6} \, .
\end{equation}

In order 
to understand the geometry of the problem one has to consider the relation
between the reference system of the detector (which we call {\it global})
and the {\it local} reference system with the origin in the center of the Earth
and the z axis passing through the detector ($z= \overline{AB}$).
The criteria to generate each event is to randomly choose an 
angle $\theta_{p}$ between $\theta^{*}$ and 0 and $\varphi_{p}$ between 0 and 
$2\pi$. This meens to extract a point $P$
on the Earth surface inside the area $A$.
The coordinates of this point in the local reference system are
($R_{\oplus} \sin\theta_{p} \cos\varphi_{p}$,
$R_{\oplus} \sin\theta_{p} \sin\varphi_{p}$, $R_{\oplus}\cos\theta_{p}$).
Then the direction of the shower is generated randomly
(the zenith angle $\theta$ and the 
azimuth $\varphi$) with respect to a reference system with the origin
in the point $P$ and the z axis at the vertical of the Earth surface.
Being Y the slanth distance from the origin $P$ 
to the first interaction point in the atmosphere $X_{0}$
(which is generated at the top of the atmosphere considering the 
interaction length of protons) the coordinates of any point $P(Y)$
along the track in this reference system are ($Y\sin\theta \cos\varphi$,
$Y\sin\theta \sin\varphi$,$Y \cos\theta$).
It is straightforward to rotate this reference system into the global one
which has the origin in the detector (B) and the z axis  
toward the center of the Earth along the vertical of the Earth 
($\overline{BA}$)
The coordinates of the point $P(Y)$ in this sistem are (x(Y), y(Y), 
z(Y)) and the plane xy is the pixel matrix plane. 
The angular coordinates of the point $P(Y)$ on the pixel matrix are
($\sin\theta_{m}\cos\varphi_{m}$, $\sin\theta_{m}\sin\varphi_{m}$) = 
$(x/D, y/D)$, where $D=\sqrt{x^{2}+y^{2}+z^{2}}$ is the distance
of $P(Y)$ from the detector.  

Assuming a total number of pixels of $N_{0} = 10^{6}$, 
located on a circular matrix, they can
be defined as squares and labelled between
$-N_{A}$ and $N_{A}$ where $N_{A} = \sqrt{\frac{N_{0}}{\pi}}$ 
($\pi N_{A}^{2} = N_{0}$) along x and y.
Hence, their constant angular width is $dx = dy = \sin\theta_{FOV}/N_{A}$.
The number of pixels which effectively view tracks
(i.e. satisfy $A_{ij}=\sqrt{(x_{i}/D)^{2}+(y_{j}/D)^{2}}< 
\sin\theta_{FOV}$) is 999289 in this arrangement
and the total solid angle seen by the pixels is 
$d\Omega = \sum_{ij} \frac{dx dy}{\sqrt{1-A^{2}_{ij}}} = 0.84$ sr.  

In Tab.~\ref{tab1} the main parameters of the mission and
the transmission and detection efficiencies \cite{OWL} are summarized.
The product of all the efficiency terms gives a reduction factor of the signal
of $\varepsilon = 0.053$. 

\begin{table} [h]   
\caption{Main characteristics of the simulated detector: half of the
field of view angle, orbit height, number of pixels, area of the mirror, 
optical efficiency of the mirror, atmospheric Rayleigh transmission,
ozone transmission and quantum efficiency of the photodetector.} 
\label{tab1}
\begin{center}       
\begin{tabular}{|c|c|c|c|c|c|c|c|c|}
\hline
\rule[-1ex]{0pt}{3.5ex}  $\theta_{FOV}$&$H$&N. of&area&optical&UV & 
atmosph.&atmosph.& Q.E. \\
\rule[-1ex]{0pt}{3.5ex}  & & pixels&of &effic.& filter& 
Rayleigh& ozone& \\
\rule[-1ex]{0pt}{3.5ex}  & &       & mirror& &effic. 
 & transm.&transm.& \\
\hline
\rule[-1ex]{0pt}{3.5ex} 
$30^{\circ}$&500 km&10$^{6}$&5 m$^{2}$&0.85   & 0.6&0.568& 
0.736&0.25 \\ 
\hline
\end{tabular}
\end{center}
\end{table}

Once a primary particle has been generated with its incident energy and 
its zenith and azimuth angles, the relation between
the slant distance along the track $Y$ (km) ($Y$ = 0 on the Earth surface) 
and the vertical height $Z$ along the vertical of the Earth 
(Z is 0 at the detector level and Z is maximum at the Earth surface) is:

\begin{equation}
\label{eq:z}
Z = \sqrt{R^{2}_{\oplus} + Y^{2} + 
2 R_{\oplus} Y \cos\theta} - R_{\oplus} \, .
\end{equation}

The slant distance is connected to the slant depth $X$ 
(gr/cm$^{2}$) along the track considering the density of the atmosphere,
which varies with height.
The track slant depth is divided into 30 equispaced pieces, whose
size is then calculated using a shower development code.
The shower cascade algorithm is based on a routine named UNICAS 
\cite{Gaisser}
which generates a shower due to the interaction of a single nucleon 
starting at a slant depth $X_{0}$ and interactions 
and decays of particles are treated. The hadronic interactions
are treated according to the splitting algorithm by Hillas 
(described in \cite{Gaisser}) with threshold energies (down to which 
particles are tracked) $\sim 89 \times 10^{3}$ TeV.
Cross sections for hadron interactions are obtained from the SIBYLL model
\cite{SIBYLL}. 
The SIBYLL model is based on the idea that the increase in cross section is 
driven by the production of minijets \cite{Cline} with particular emphasis
on the fragmentation region (appropriate for cosmic rays) and on collisions
of hadrons with light nuclei. The ideas of the dual parton model 
\cite{Capella}
are incorporated. The code is tailored and
efficient at energies up to at least $10^{20}$ eV.
The main outcomes of the code are the depth and size of shower maximum.  

\begin{figure}[t]
   \begin{center}
   \begin{tabular}{cc}
\psfig{figure=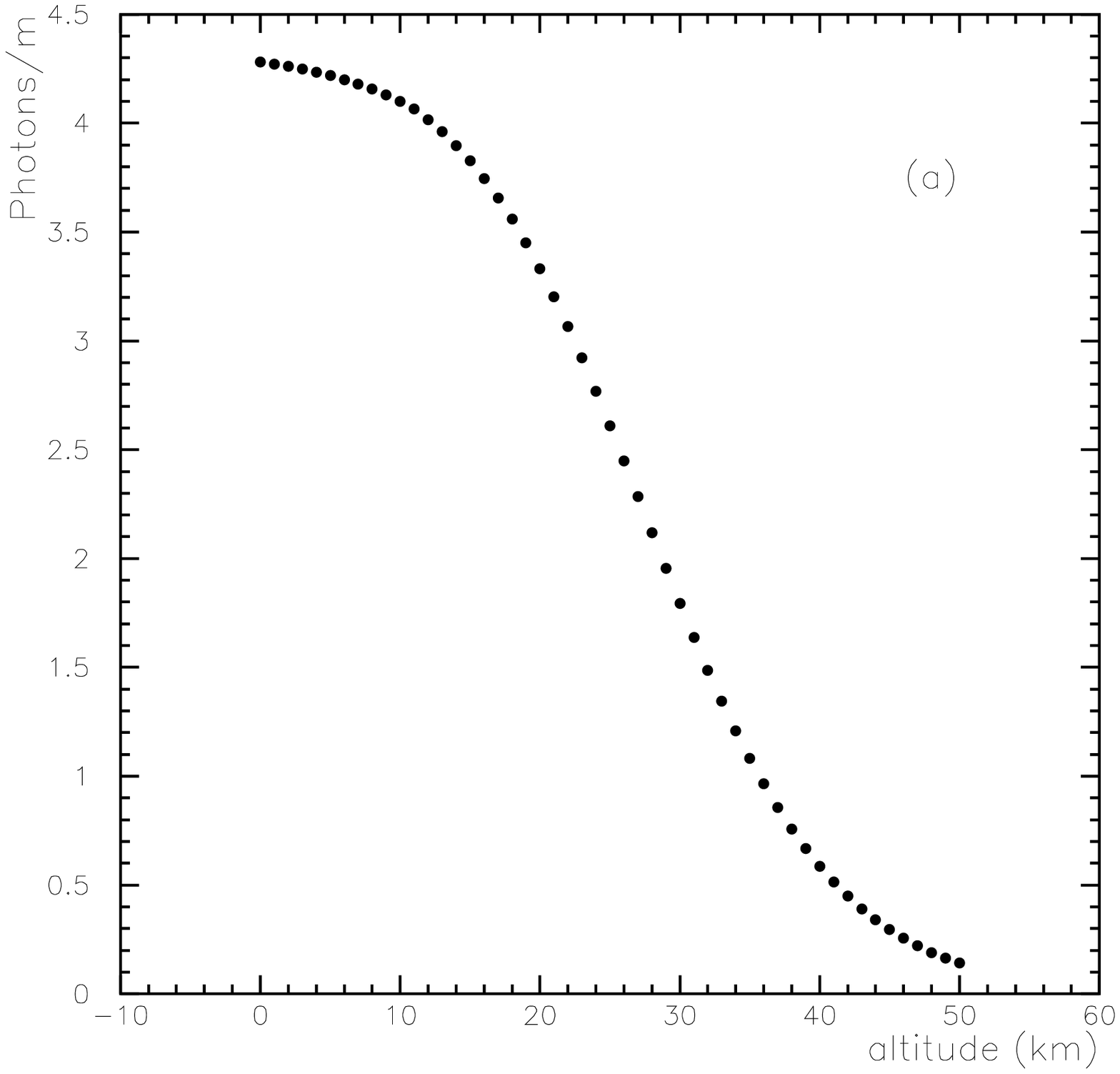,width=7.cm,height=7.cm}&
\psfig{figure=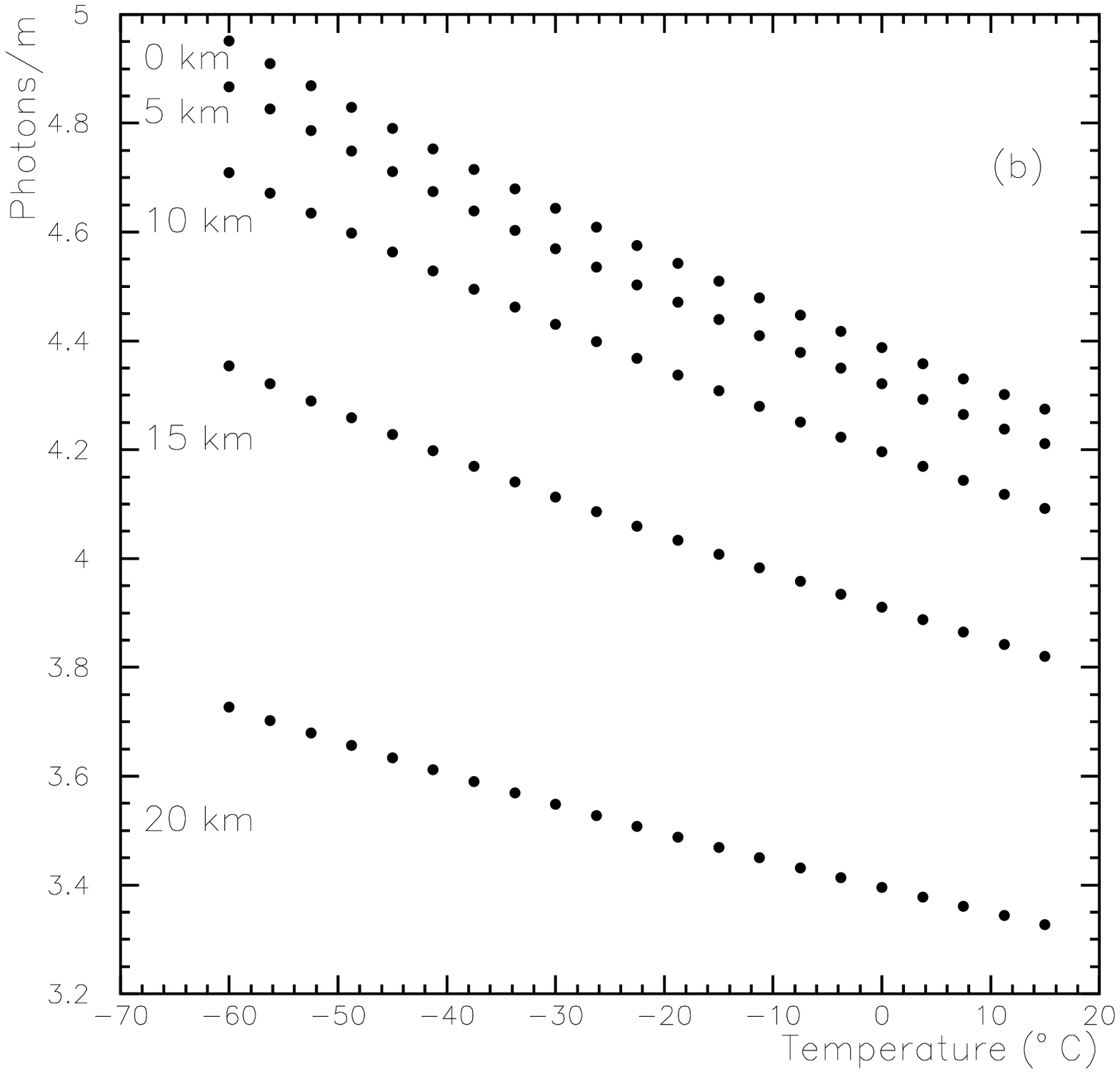,width=7.cm,height=7.cm}
   \end{tabular}
   \end{center}
   \caption[Fluorescent yield] 


{ \label{eff}	  

(a) Fluorescent yield of photons/particle/meter
in the range 300-400 nm as a function of the altitude at STP. 
(b) Fluorescent yield 
($\gamma/particle/m$) as a function of temperature for 5 different
altitudes \protect\cite{Dai}.} 

\end{figure}

Once the number of charged particles for each step and  
the shower maximum and size at maximum have been determined,
a track of photons is produced on the detector focal plane.
The photoelectron number per pixel obtained for each step
is extracted from a poissonian
distribution of mean value:

\begin{equation}
\label{eq:gamma}
(N_{p.e.}/pixel)_{i} = \frac{A_{mirror}}{4 \pi R^{2}} 
\varepsilon_{light} S \varepsilon \, , 
\end{equation} 

where $i$ labels the piece of track,
$A_{mirror}$ is the area of the collecting mirror, $R$ is the
distance of the piece from the detector, 
$\varepsilon_{light}$ is the number of photons produced per particle
per unit length,
$S$ is the size of each piece and $\varepsilon$ is the efficiency 
factor which includes all terms in Tab.~\ref{tab1}. 
In Fig.~\ref{eff} the behaviour of $\varepsilon_{light}$ as a function
of the altitude and of the temperature is shown \cite{Dai}.
The fluorescent yield at STP in the range 300-400 nm is 4.27 
photon/meter/particle.

The most interesting information on simulated events are stored, namely 
the energy and the zenith angle of the incoming particle, 
the number of photoelectrons per pixel on the detector, 
the number of photoelectrons in pixels which detect a signal 
(above the detection threshold of 1 p.e.),
the total number of pixels hit and those above threshold, the size
and the slant depth of the maximum of the shower.

In Fig.~\ref{ev} we show two simulated ``photon tracks'' on the pixel 
matrix for two values of the energy of the primary particles
differing by 1 order of magnitude and impinging with the same zenith angle.

\begin{figure}[t]
\begin{center}
   \begin{tabular}{cc}
\psfig{figure=19_ev.ps,width=9.cm,height=10.cm}
\psfig{figure=20_ev.ps,width=9.cm,height=10.cm}

   \end{tabular}
\end{center}
\caption[Schematic plot of one event] 


{\label{ev}	  

Schematic plots of events as seen on the pixel matrix (the plane 
($\sin\theta_{m}\cos\varphi_{m}$, $\sin\theta_{m}\sin\varphi_{m}$)):
the event on the left (right) has been generated by a primary particle 
of energy 11 EeV (111 EeV). The zenith
angle of  the primary is in both cases $76.2^{\circ}$.
The letter A stands for the first interaction point. 
The shown track represents the projection of the atmospheric track 
onto the matrix plane. The squares are the pixels with at least 1 p.e.. 
There are 27 (78) pixels hit for a total number of 35 (465) photoelectrons
for the event on the left (right).} 
\end{figure}

\begin{figure}[t]
   \begin{center}
   \begin{tabular}{cc}
\psfig{figure=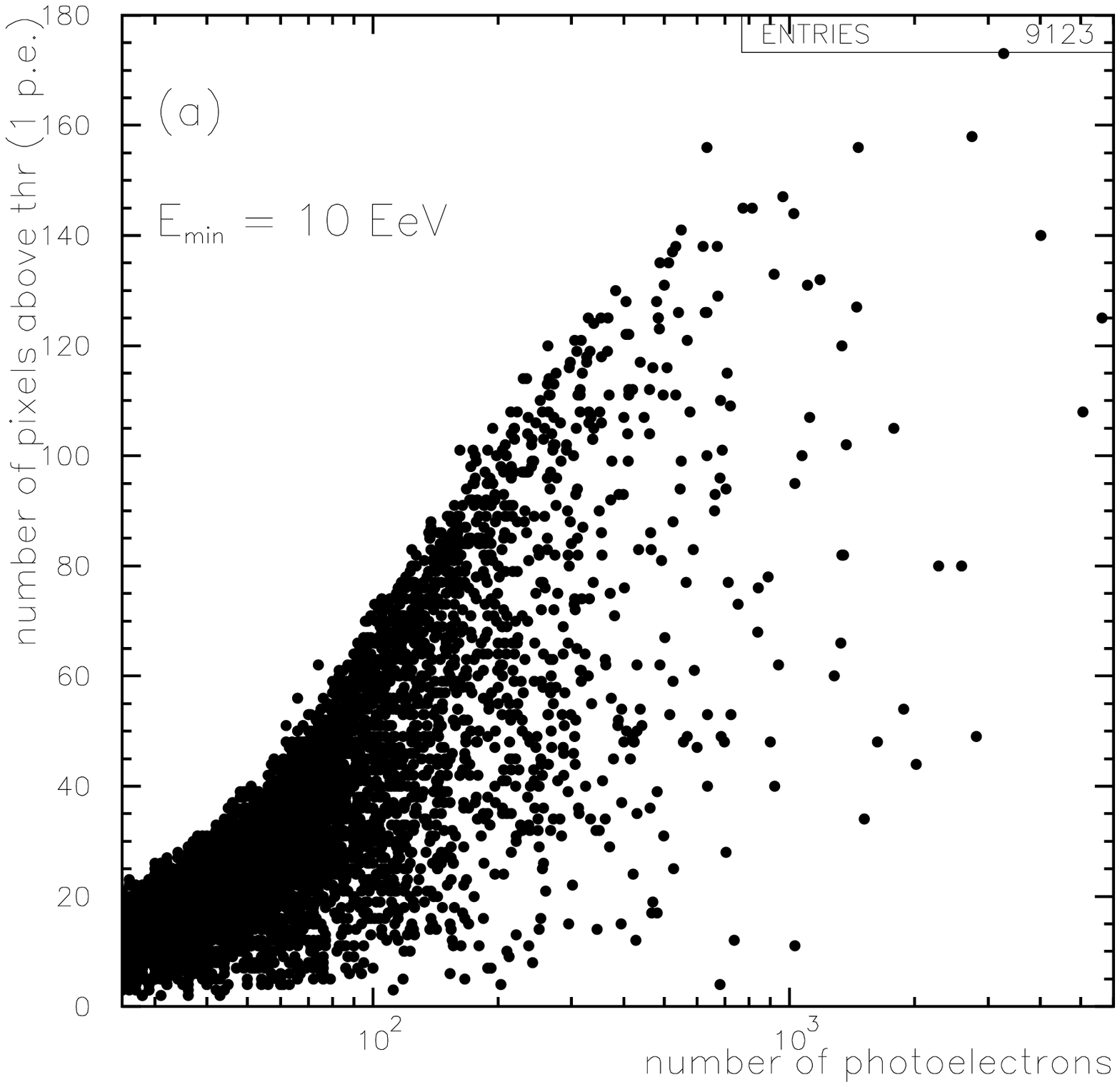,width=7.cm,height=7.cm}&
\psfig{figure=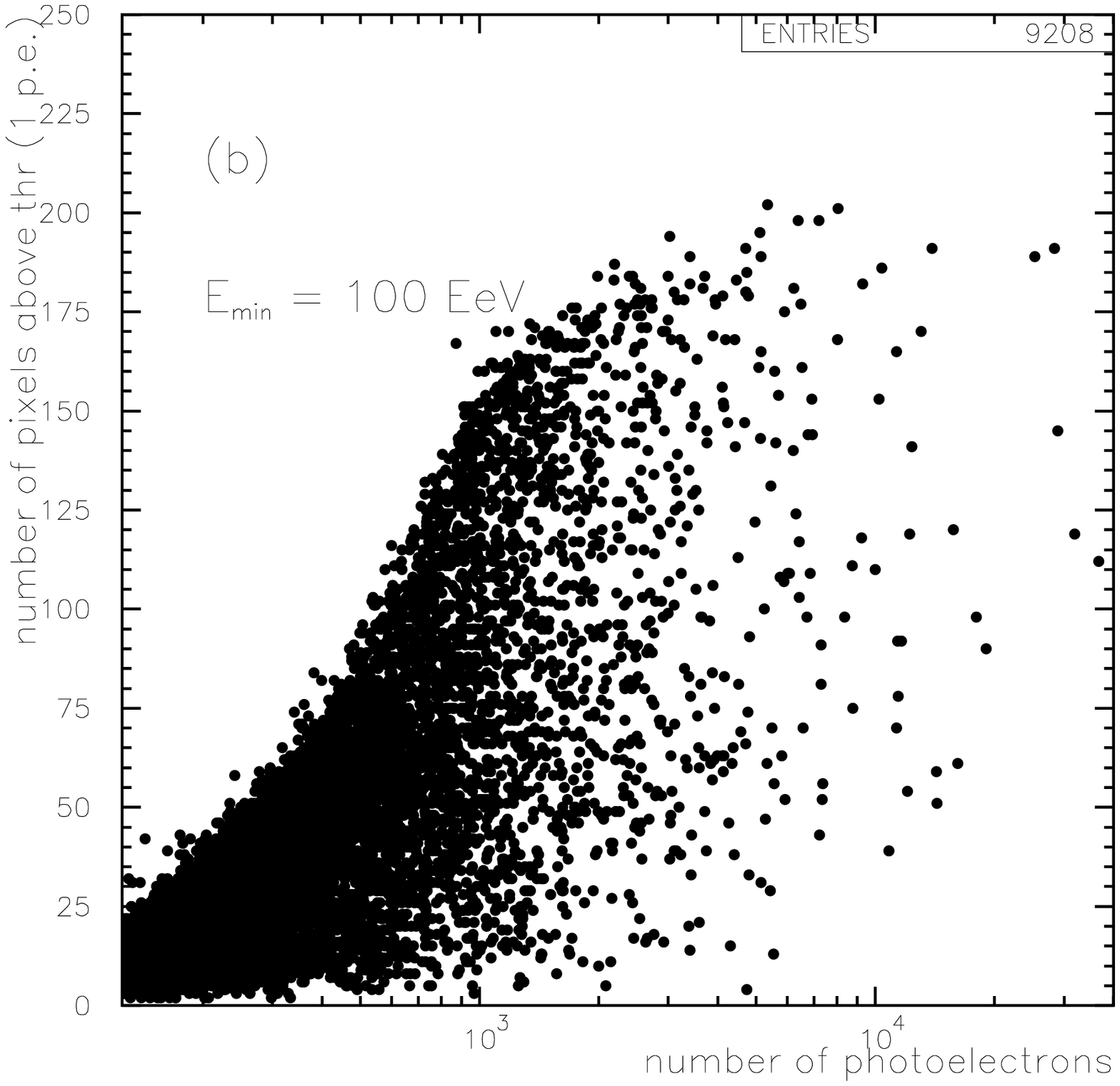,width=7.cm,height=7.cm}
   \end{tabular}
   \end{center}
   \caption[Number of photoelectrons vs number of pixels] 


{\label{size}	  

Number of pixels hit above
threshold (at least 1 p.e.) vs number of photoelectrons  
(a) for incident primaries with energy above 10 EeV; 
(b) above 10$^{2}$ EeV. 10000 events were generated for each plot. 
Those which produce at least 1 p.e. on the pixels are 9123 in the
case of 10 EeV and 9208 in the case of 100 EeV. In order to match the
size of pixels, the number of steps in which the track is divided in the 
atmosphere (30) is increased to 300 (maximum number of pixels hit 
in one event) when projecting the track onto the pixel matrix.} 

\end{figure}

\begin{figure}[t]
   \begin{center}
   \begin{tabular}{c}
\psfig{figure=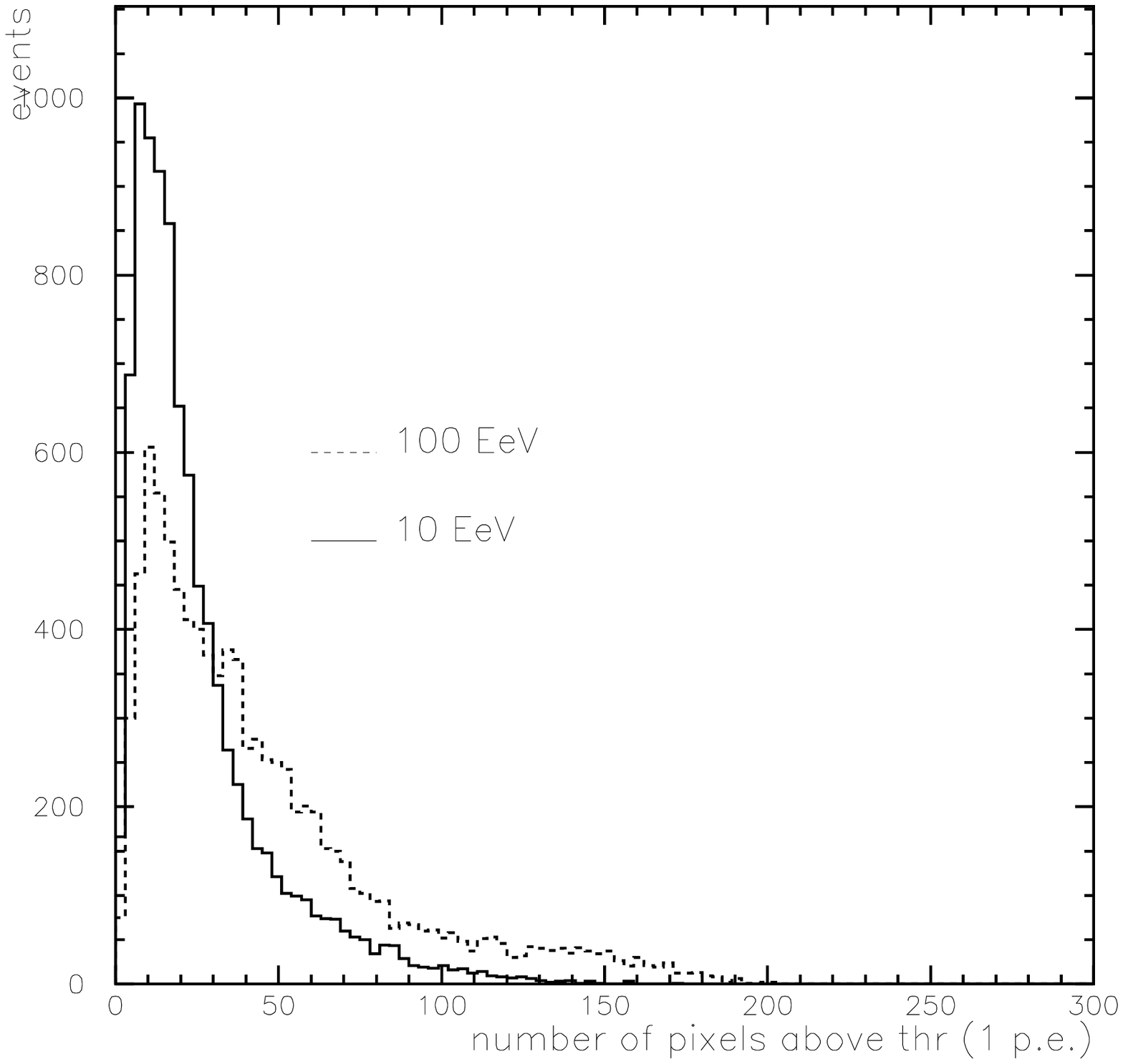,width=10.cm,height=7.cm}
   \end{tabular}
   \end{center}
   \caption[number of pixels] 


{\label{pixel}	  

Distribution of the number of pixels hit above
threshold (at least 1 p.e.) for incident primaries with energy above 10 EeV 
and above 10$^{2}$ EeV.
} 

\end{figure}

In Fig.~\ref{size} the total number of pixels above threshold 
versus the total number of photoelectrons  
is shown for all events for minimum energies of the primary spectrum 
$E_{min} = 10$ and 100 EeV.
In Fig.~\ref{pixel} the distribution of the number of pixel hit
above threshold for all the events is shown.
The mean value of the number of detected photons is increased 
by a factor of 2 when the minimum energy is increased from 10 EeV
to 100 EeV.

\begin{figure}[t]
   \begin{center}
   \begin{tabular}{cc}
\psfig{figure=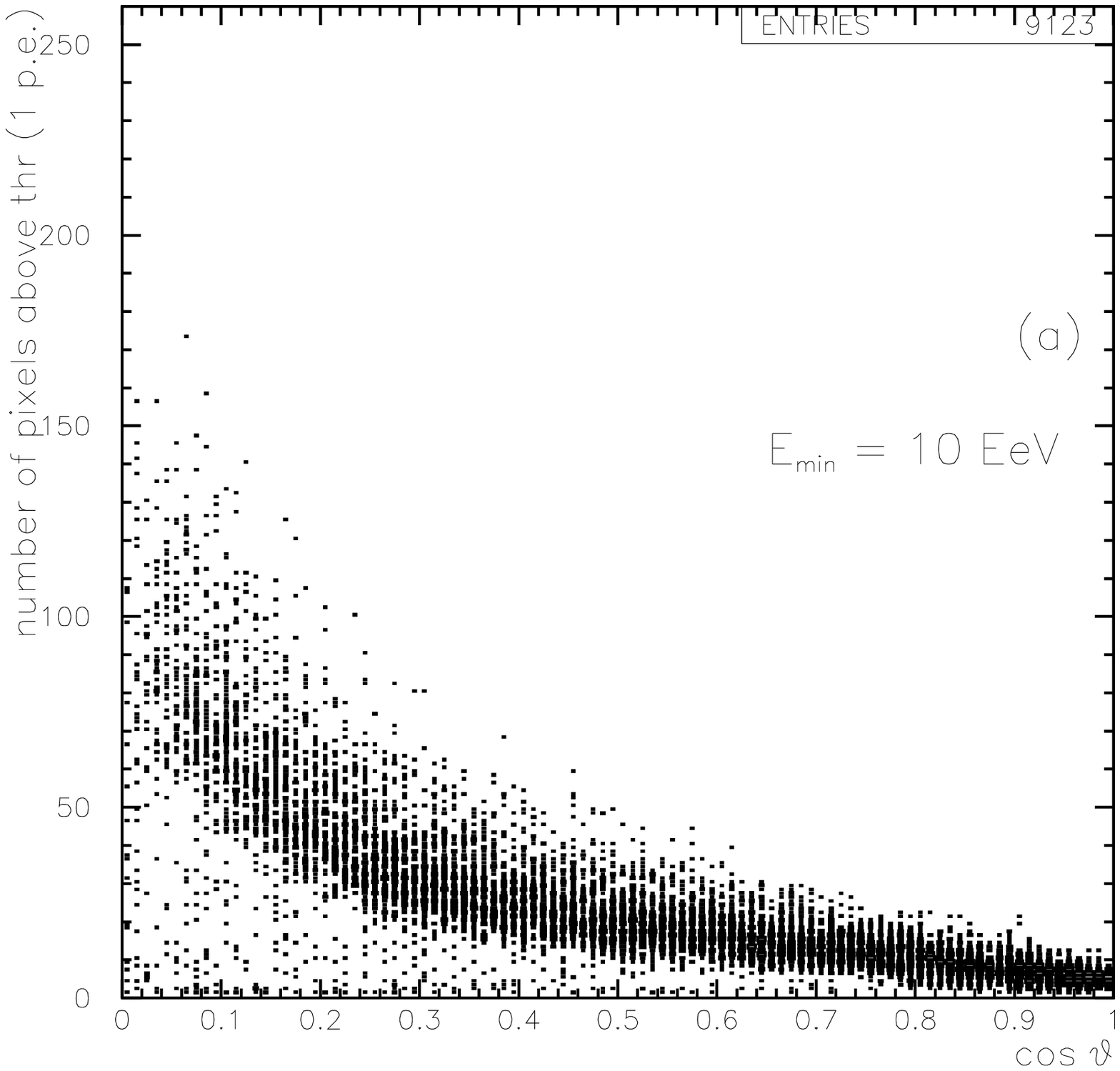,width=7.cm,height=7.cm}&
\psfig{figure=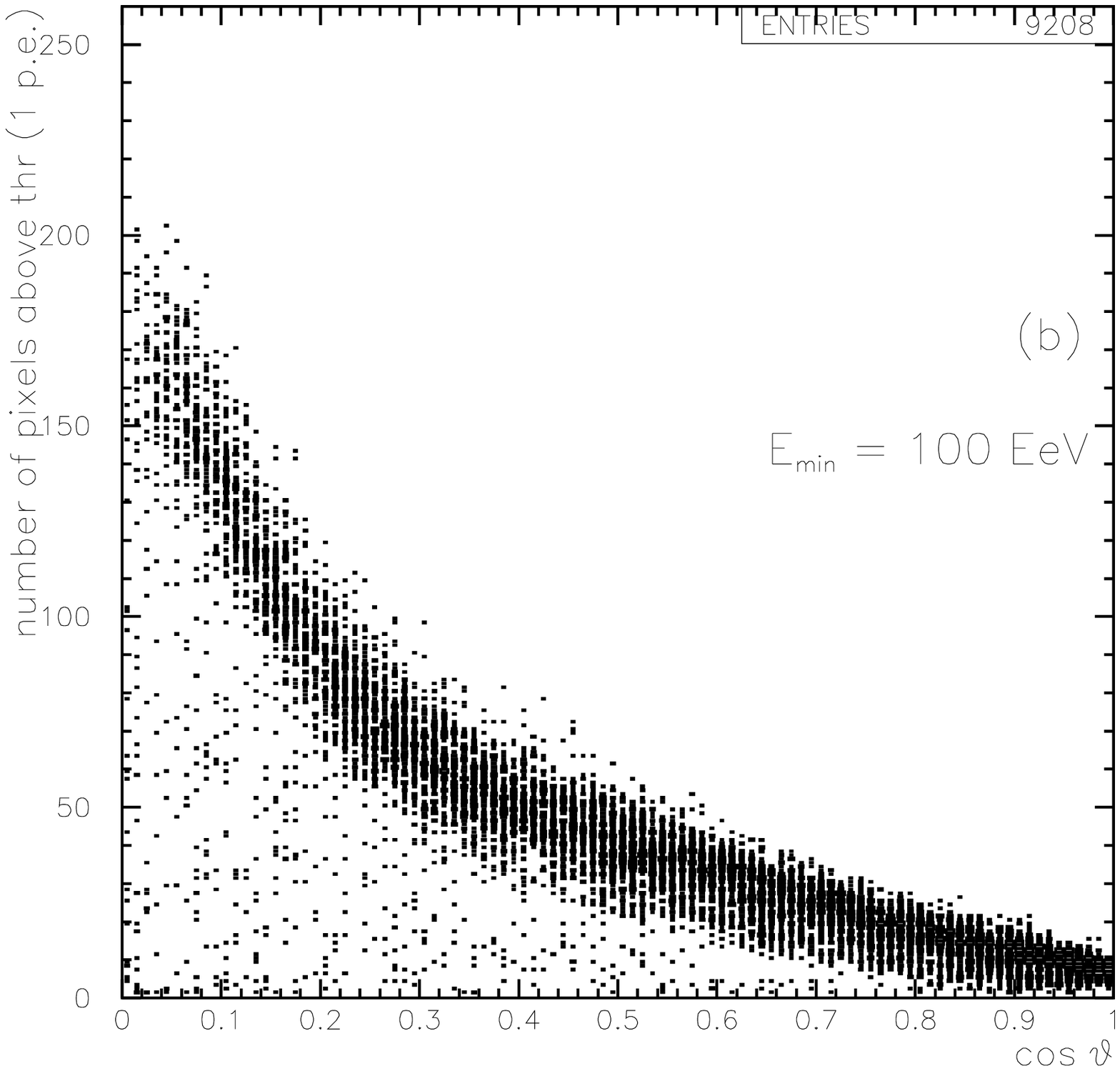,width=7.cm,height=7.cm}
   \end{tabular}
   \end{center}
   \caption[Number of pixels] 


{ \label{np}	  

(a) Number of pixels which measure at least one 
p.e. as a function of the cosine of the zenith angle of the incident primaries
with energy above 10 EeV. 
(b) Same as (a) but for minimum energy $10^{2}$ EeV. The two plots
have been obtained generating 10$^{4}$ events.} 

\end{figure}

In Fig.~\ref{np} the number of pixels hit above threshold summing over all 
events is shown for two 
values of the minimum energy considered (10 and 10$^{2}$ EeV)
versus the cosine of the zenith angle of the track in the
local system.
These plots are very important in establishing two facts:
\begin{enumerate}
\item even if vertical tracks are more abundant than horizontal ones,
it is clear that the detection of almost horizontal tracks is favoured 
by the large numer of pixels hit. This is important in view of the
detection of neutrinos as horizontal showers crossing huge
amounts of atmosphere;
\item the energy threshold of the detector is a critical parameter:
the aim of the mission is to investigate the region of the ankle of the
cosmic ray spectra, but the possibility of measuring showers with
energy as low as 10$^{18}$-10$^{19}$ eV
could provide interesting hints for what concerns neutrino physics 
items such as investigations on active galactic nuclei and neutrino
astronomy. Moreover the lower the threshold the higher the statistics that can
be achieved due to the power law behaviour of the cosmic ray spectrum.
Another important aspect of having a low threshold is that the
space mission data could be compared to the ground based ones, hence
providing a calibration opportunity for this new kind of detector with
respect to a more traditional technique.
The plot in Fig.~\ref{np} (a) shows that the detection at 10$^{19}$ eV
is feasible.
\end{enumerate}

\section{CONCLUSIONS}
At the current stage of investigation presented in this work 
the simulation code
is already an useful tool for designing Airwatch type missions.
As a further step, the timing of each photon
reaching the focal plane will be implemented. This will give the  
development in time of the detector electronic signals. 
This knowledge is fundamental in order to derive the shower 
axis direction in the Airwatch concept. 
This will allow an estimation of the background level, which depends on the
integration time of the acquisition system.

\acknowledgments
We would like to thank T. K. Gaisser for useful discussions and
contributions to this work. Thanks are due also to the whole 
Airwatch Collaboration, and in particular
John Linsley and Livio Scarsi for useful discussions on the
Airwatch idea.

\bibliography{spie}   
\bibliographystyle{spiebib}   

\end{document}